\documentclass[doublecol,figures]{epl2} 
\usepackage{amsmath,bbm}
\usepackage{subfigure}
\usepackage[colorlinks]{hyperref}
\hypersetup{linkcolor=[rgb]{0.2,0.4,0.7}} 

\newcommand{\pd}{\partial}
\newcommand{\bs}{\boldsymbol}

\renewcommand{\i}{\mathrm{i}}
\let\oldpi\pi
\renewcommand{\pi}{\mathrm{\oldpi}}
\newcommand{\e}{\mathrm{e}}

\renewcommand{\revision}{}

\title{Spin effects in electron vortex states}

\author{Ruben Van Boxem\inst{1}\thanks{E-mail: \email{ruben.vanboxem@ua.ac.be}} \and Jo Verbeeck\inst{1} \and Bart Partoens\inst{1}}
\shortauthor{Ruben Van Boxem \etal}

\institute{                    
  \inst{1} EMAT \& CMT, University of Antwerp - Groenenborgerlaan 171, 2020 Antwerp, Belgium
}

\abstract
{
  The recent \revision{experimental realization} of \revision{electron vortex beams} opens up a wide research domain previously unexplored.
  The present paper explores the relativistic properties of these \revision{electron vortex} beams, and quantifies deviations from scalar wave theory.
  It is common in electron optics to use the Schr\"odinger equation neglecting spin.
  The present paper investigates the role of spin and the total angular momentum $J_z$ and how it pertains to the vortex states.
  As an application, we also investigate if it is possible to use holographic reconstruction to create novel total angular momentum \revision{eigenstates in} a Transmission Electron \revision{Microscope.}
  It is demonstrated that relativistic \revision{spin coupling} effects disappear in the paraxial limit, and spin effects in holographically created electron vortex beams \revision{can only be exploited by using specialized magnetic apertures}.
}

\begin{document}

\maketitle

\section{Introduction}

Singular optics is the study of light waves with a phase singularity~\cite{Nye}.
For two decades, since the experimental discovery of optical vortices~\cite{Heckenberg}, study of singular properties in the scalar theory of light has led to fascinating applications in various areas of physics~\cite{ONeill,Padgett,He,Thide,Berkhout,Swartzlander}.
Much more recently, the concepts and experimental setups from singular optics found their way to electron microscopes~\cite{Allen,Uchida,Verbeeck}.
Different methods are being investigated for the production of electron vortices~\cite{Verbeeck_spiral,Schattschneider_fork} and possible applications at the nanoscale are surfacing~\cite{Verbeeck_tweezers}.
The first steps into developing a theory of electron vortices have been taken~\cite{Schattschneider_theory,Bliokh_semiclassical,Bliokh_magnetic}.
The quantum mechanical description of electron optics coincides for the most part with scalar optics \revision{due to the similarity of their respective wave equations.}
\\
Both photon polarization and electron spin determine a spin angular momentum vector, which contributes together with the orbital angular momentum vector to the total angular \revision{momentum vector}: $\bs J = \bs L + \bs S$.
In this paper, the cylindrically symmetric case is studied, and the quantities of interest are the projections of the angular momenta on the $z$-axis (parallel to beam propagation): $J_z = L_z + S_z$.
In what follows, cylindrical symmetry is assumed, and all references to ``angular momentum" refer to the projected angular momenta on the vertical $z$-axis.
One can pose the question of the influence of the bosonic nature of the photon versus the fermionic character of the electron.
Electrons have half-integer spin, which translates to half-integer \revision{total angular momentum} as a proper observable, where photons have integer \revision{spin angular momentum}, and thus integer \revision{total angular momentum}.
The difference in \revision{total angular momentum} is elucidated by inspecting the Dirac equation, for which $L_z$ is not a good quantum number, and it is only the \revision{total angular momentum} $J_z$ that commutes with the Dirac \revision{Hamiltonian.
Yet} experiments show that electron vortex beams behave as their scalar counterparts, and \textit{e.g.} spin seems to have little to no influence in the entire \revision{Transmission Electron Microscopy (TEM)} image formation process~\cite{Rother,Fujiwara,Ferwerda,Jagannathan}, \revision{in the acceleration voltage range of interest (100-500\;keV)}.
This is caused by the relatively small magnitude of the spin effects (or conversely, the high electromagnetic fields required to show the effects), and here it is shown that in the paraxial limit one can speak of proper $L_z$ eigenstates even in Dirac theory.
\\
\revision{The availability of two spin components does lead} to the possibility of distinct $J_z$ eigenstates\cite{Leary}, which possess interesting properties such as a skyrmionic spin distribution~\cite{Wang}.
\revision{The general conclusions of what is described here are valid for any fermion, and can be applied equally well to neutron beams.}
We pose the question whether such $J_z$ states can be produced in similar ways as the $L_z$ eigenstates, \textit{e.g.} using holographic masks~\cite{Verbeeck,Uchida}.

In what follows, natural units are used: $\hbar=c=1$.
Operator expectation values are denoted $\langle O \rangle$. When the state is an eigenstate, its values are denoted without brackets as $O$.

\section{Phase singularities and electron vortices}

The simplest description (using a scalar wave theory) of a phase singularity in the paraxial regime in a transverse plane can be written down in terms of the wave function (or in \revision{scalar} optics, \revision{a component of} the electric field $\bs E$), with a singular phase factor:
\begin{equation}
\Psi(r,\varphi) = \e^{\i m \varphi} \psi(r).
\end{equation}
One can readily see that this is an OAM eigenstate: $L_z \Psi = -\i\pd_\varphi \Psi = m \Psi$.
The phase increases linearly with $\varphi$, resulting in a singular point at $r=0$ in the center of the circular phase ramp, where the phase is undefined.
Experimental production of these types of wave functions can be accomplished by using computer-generated holograms~\cite{Verbeeck}.
These binary masks can be calculated from the interference of a tilted plane wave with the target vortex mode \revision{(see below)}.
Using the principles of holographic reconstruction, illuminating the mask with the reference beam generates the target wave as output.
In the case of binary masks, one also gets several higher diffraction orders in the far field.
This scalar optics technique allows to create any OAM and even superpositions of $L_z$ are \revision{possible~\cite{Guzzinati}.}

\revision{The quantum mechanical properties of two component electron vortex beams will now be investigated using the Pauli equation.}

\section{OAM Pauli spinors}
\subsection{Mathematical solutions}

The non-relativistic equation of choice to describe electron polarization is the 2-component Pauli equation.
It provides a non-relativistic description of spin and its interaction with electromagnetic fields, which describes the processes in an electron microscope sufficiently.
The (stationary) Pauli equation (assuming $\psi \propto \Psi(\bs r) e^{-\i Et}$) with an electromagnetic field can be written as follows,
\begin{equation} \label{eq:pauli_equation}
\left( \frac{1}{2m} \bs \sigma \cdot (\bs p - e \bs A)^2 + e \Phi + \mu_B \bs \sigma \cdot \bs B\right) \Psi = E \Psi,
\end{equation}
with vector potential $A=(\Phi,\bs A)$ representing an external electromagnetic field, $\bs B = \nabla\times\bs A$, $\phi$ the scalar (electrostatic) potential, $\bs \sigma$ the Pauli vector, $e$ the electron's charge, $\mu_B=\frac{e}{2m}$ the Bohr magneton, and $\Psi=\begin{pmatrix} \psi^1 \\ \psi^2\end{pmatrix}$ the 2-component Pauli spinor.
In cylindrical symmetry expressed in the coordinates $(\rho,\varphi,z)$, \revision{in the absence of electromagnetic fields}, the solutions are readily given by
\begin{align} \label{eq:pauli_vortex_spin}
\Psi_n^+(\rho,\varphi,z) &= \e^{\i k_z z}\begin{pmatrix} \e^{\i n\varphi}J_n(\kappa \rho) \\ 0 \end{pmatrix}, \notag\\
\Psi_n^-(\rho,\varphi,z) &= \e^{\i k_z z} \begin{pmatrix} 0 \\ \e^{\i n\varphi} J_n(\kappa \rho) \end{pmatrix}.
\end{align}
These states are eigenstates of the energy $E$, the forward momentum $k_z$, the transverse momentum $\kappa$, the OAM $L_z=n$, and the spin $S_z$ (labeled with the $\pm$ superscript), and thus also of $J_z=L_z+S_z$.
The radial functions $J_n(x)$ are the $n$-th order cylindrical Bessel functions of the first kind.
\revision{These states are orthogonal basis functions that can be used to build up any beam shape, including those emerging from a holographic mask (see below).}

\subsection{\revision{Angular momentum analysis}}

In principle, one can combine any $\Psi_n^+$ with any other $\Psi_m^-$ and still have a valid solution, although in general it won't be an $S_z$ or $L_z$ eigenstate anymore.
The case where $m=n+1$ is interesting because this combination is also an eigenstate of the TAM $J_z=n+\frac{1}{2}$:
\begin{equation} \label{eq:pauli_vortex_j}
\Psi_n^{h_\perp=\pm1}(\rho,\varphi,z) = \e^{\i k_z z} \begin{pmatrix} \e^{\i n\varphi} J_n(\kappa \rho) \\ \pm \e^{\i (n+1) \varphi} J_{n+1}(\kappa \rho) \end{pmatrix}.
\end{equation}
This state has traded \revision{the quantum numbers} $L_z$ and $S_z$ for $J_z$ and the non-relativistic limit of the transverse helicity $h_\perp = \sigma\cdot \frac{\bs p \times \bs e_z}{|\bs p \times \bs e_z|}$.
These states are not OAM eigenstates, and the two scalar spin components have a different OAM.
This state's \revision{expectation value $\langle L_z \rangle$} is equal to its $J_z$ eigenvalue:
\begin{equation} \label{eq:pauli_spinor_jz_oam}
\langle L_z \rangle = \frac{1}{2}[n+(n+1)] = n+\frac{1}{2} = J_z,
\end{equation}
This should come as no surprise because $\langle J_z \rangle = \langle L_z \rangle + \langle S_z \rangle$ and $\langle S_z \rangle$ is equal to $0$ for a spin unpolarized \revision{beam.

A} general Pauli electron vortex beam can be parametrized with $\alpha\in\mathbbm{R}$ (the radial functions $f(r)$ and $g(r)$ are assumed to be normalized to 1):
\begin{equation} \label{eq:pauli_general_spinor}
\Psi(r,\varphi,z) = \e^{\i k_z z} \begin{pmatrix} \frac{1}{\sqrt{1+\alpha^2}} \e^{\i n\varphi} f(r) \\ \frac{\alpha}{\sqrt{1+\alpha^2}} \e^{\i n^\prime\varphi} g(r) \end{pmatrix},
\end{equation}
where $n$ and $n^\prime$ are the two spin components' OAM.
This normalized state ($|\Psi|^2=1$) is an exact solution of the field-free Pauli equation, but not (in general) an eigenstate of any angular momentum \revision{operator}.
\\
If the weight of both components is equal ($\alpha=1$), one has
\begin{equation}
\langle L_z \rangle = \frac{1}{2}(n+n^\prime),
\end{equation}
which leads to an $L_z=n$ eigenstate if $n=n^\prime$, and
\begin{equation}
\langle J_z \rangle = \frac{1}{2}[(n+1/2)+(n^\prime-1/2)] = \frac{1}{2}(n+n^\prime),
\end{equation}
\revision{resulting in} $\langle J_z \rangle = n$ if $n=n^\prime$.
So in the case of identical vortices in both spin components, we have $L_z = \langle J_z \rangle = n$.
\revision{Note that this is the inverse of \eqref{eq:pauli_spinor_jz_oam}, where the $J_z$ eigenvalue is equal to the expectation value $\langle L_z \rangle$.}
\revision{Here it is the $L_z$ eigenvalue that is equal to the expectation value $\langle J_z \rangle$.}
\\
If the weight of both components is not equal, one has for the angular momenta expectation values:
\begin{subequations}
  \begin{align}
  \langle L_z \rangle &= \frac{n}{1+\alpha^2} + \frac{\alpha^2 n^\prime}{1+\alpha^2} = \frac{n+\alpha^2 n^\prime}{1+\alpha^2}, \\
  \langle S_z \rangle &= \frac{\frac{1}{2}}{1+\alpha^2} + \frac{-\frac{1}{2}}{1+\alpha^2} = \frac{1}{2} \frac{1-\alpha^2}{1+\alpha^2}, \\
  \langle J_z \rangle &= \langle L_z + S_z \rangle = \frac{n+\alpha^2 n^\prime}{1+\alpha^2} + \frac{1}{2}\frac{1-\alpha^2}{1+\alpha^2}.
  \end{align}
\end{subequations}
\\
In the special case $n^\prime=n+1$, spinor \eqref{eq:pauli_general_spinor} is an eigenstate of $J_z$ with eigenvalue $n+\frac{1}{2}$, and this is unaffected by the value of $\alpha$.
\revision{Even though} only $J_z$ is a proper quantum number, for the special case of \revision{spin} unpolarized electrons ($\alpha=1$), the $J_z$ eigenstate also has $\langle L_z \rangle = J_z \revision{ = n+\frac{1}{2}}$.

\section{\revision{Holographic masks and spin}}
\subsection{Selection on $L_z$ by a holographic mask} \label{sec:pauli_hologram}

\begin{figure}
\centering{}
\includegraphics[width=\linewidth]{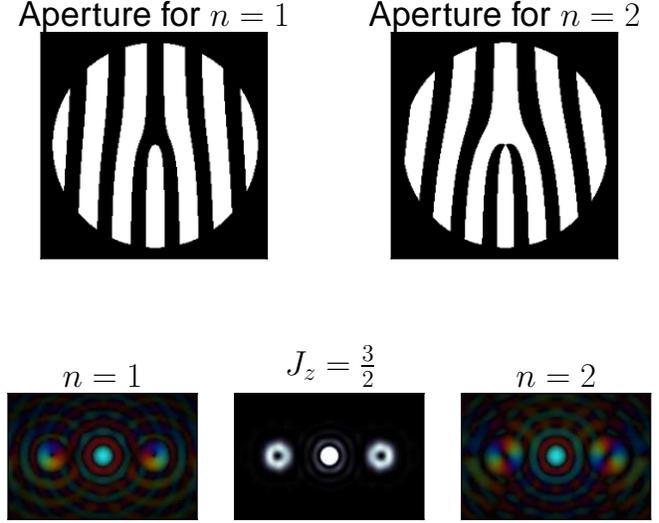}
  \caption{The \revision{simulated} farfield of a scalar $n=1$, $n=2$, and a spinor $J_z=\frac{3}{2}$ (equation \eqref{eq:pauli_vortex_j}) with the radial distribution caused by a circular aperture with a fork pattern. \revision{The colour signifies the phase.} The doughnut shapes are phase vortices, with a phase ramp of $0$ to $n2\pi$, and a central zero-density phase singularity. \label{fig:spinor}}
\end{figure}

\revision{
Production of electron vortices is most easily accomplished by using the optical technique called \emph{Fourier transform holography}, in which the recorded interference pattern of two beams allows reconstruction of one of the beams by illumination of the interference hologram by the other beam.
\\
Mathematically, this comes down to a simple formalism.
Let $\Psi_\mathrm{R}$ be a reference wave (\textit{e.g.} (tilted) plane wave) and $\Psi_\mathrm{T}$ the ``target" wave (\textit{e.g.} a vortex), the tilde and $\mathcal{F}$ denote a Fourier transform, and the star product denotes convolution:
\begin{subequations} \label{eq:fourier_holography}
\begin{align}
\widetilde{M} &= \left|\widetilde{\Psi}_\mathrm{R} + \widetilde{\Psi}_\mathrm{T}\right|^2 \notag \\
&= |\widetilde{\Psi}_\mathrm{R}|^2 + |\widetilde{\Psi}_\mathrm{T}|^2 + \widetilde{\Psi}_\mathrm{R}^* \widetilde{\Psi}_\mathrm{T} + \widetilde{\Psi}_\mathrm{T}^* \widetilde{\Psi}_\mathrm{R} \notag \\
&= \widetilde{I}_\mathrm{R} + \widetilde{I}_\mathrm{T} + \widetilde{\Psi}_\mathrm{R}^* \widetilde{\Psi}_\mathrm{T} + \widetilde{\Psi}_\mathrm{T}^* \widetilde{\Psi}_\mathrm{R}, \\
\mathcal{F}[\widetilde{M}\widetilde{\Psi}_\mathrm{R}] &= M*\Psi_\mathrm{R} \notag \\
&= (I_\mathrm{R}^2 + I_\mathrm{T}^2) * \Psi_\mathrm{R} + |\Psi_\mathrm{R}|^2 * \Psi_T + \Psi_T^* * \Psi_\mathrm{R}^2. \label{eq:scalar_hologram_reconstruction}
\end{align}
\end{subequations}
This shows that by illuminating the interference mask by a tilted plane wave ($\widetilde{\Psi}_\mathrm{R} = \e^{\i k_x x}$), one ends up with the far field of the target wave, the reference wave, and target wave's complex conjugate displaced in the $x$ direction proportionally to $0$, $k_x$, and $2k_x$ respectively.
Fig. \ref{fig:spinor} shows the (binarized) masks for a first and second order vortex with a tilted plane wave and their far field wave function.
}
The \revision{central} far field image shows the density of a spinor $J_z$ eigenstate \eqref{eq:pauli_vortex_j}, which is the sum of the other two images.
It contains two superimposed and independent scalar vortices of different order.

\subsection{Selection on $J_z$ by a holographic mask}

In order to create a $J_z$ eigenstate such as those in \eqref{eq:pauli_vortex_j}, one needs a superposition of different OAM for each spin component.
For photons, subwavelength gratings can be used to produce such states~\cite{Gorodetski}.
This is difficult for electrons in an electron microscope, as the wavelength of an accelerated electron is typically of the order of pm.
Creating half-integer holographic masks does not have the desired effect \revision{as the mask does not contain the target spinor information.}
As previously shown by Berry~\cite{Berry}, this leads to non-integral OAM which is nothing but the sum of integral OAM contributions, which show in the wave pattern as the interference of the different OAM \revision{modes.}
\revision{A simple superposition composition of two spin-polarized vortex beams can be used instead, which is generated by an inverse OAM and spin analyzer as described in~\cite{Leach}.}

\revision{The challenge is acquiring a space-variant polarization as described by Wang \textit{et al.}~\cite{Wang} for the target states.
Modification of equation \eqref{eq:fourier_holography} by making the $\Psi$'s 2-component spinors and $M$ a matrix operator, one has instead of equation \eqref{eq:scalar_hologram_reconstruction}, a spinor form:
\begin{equation}
\mathcal{F}[\widetilde{M}\widetilde{\Psi}_\mathrm{R}] = \left[ (\Psi_\mathrm{R}+\Psi_\mathrm{T})^2 + \Psi_\mathrm{R}^\dagger \Psi_\mathrm{T} + \Psi_\mathrm{T} \Psi_\mathrm{R} \right] * \Psi_\mathrm{R},
\end{equation}
where the dagger superscript represents Hermitian conjugation so that $\Psi_\mathrm{R}^\dagger \Psi_\mathrm{T}$ and $\Psi_\mathrm{T}^\dagger \Psi_\mathrm{R}$ represent scalars, not leading to any spinor $\Psi_\mathrm{T}$. 

Approaching this problem backwards one could deduce a heuristic form of $M$.
If $\widetilde{M}\widetilde{\Psi}_\mathrm{R} = C_1 \widetilde{\Psi}_\mathrm{R} + C_2 \widetilde{\Psi}_\mathrm{T} + C_3 \widetilde{\Psi}_\mathrm{T}^*$, then
\begin{equation}
\begin{pmatrix} a & b \\ c & d \end{pmatrix} \begin{pmatrix} \psi^1 \\ \psi^2 \end{pmatrix} = C_1 \begin{pmatrix} \psi^1 \\ \psi^2 \end{pmatrix} +  C_2 \begin{pmatrix} \psi_T^1 \\ \psi_T^2 \end{pmatrix} + C_3 \begin{pmatrix} \left(\psi_T^1\right)^* \\ \left(\psi_T^2\right)^* \end{pmatrix}.
\end{equation}
Here, $C_i \in \mathbbm{C}$ and in principle one of the two last two spinors can have their up and down components switched, leading to at least one $J_z$ eigenstate.
The matrix operator $M$ can be constructed using the last term in the Pauli equation \eqref{eq:pauli_equation}: $\mu_B \bs \sigma \cdot \bs B$.
By calculating a magnetic field vector hologram in addition to a scalar (binary) aperture, one can in principle arbitrarily manipulate the spin components into $J_z$ eigenstate vortices.
This type of magnetized aperture diffraction has been considered previously in light of the transverse Stern-Gerlach effect~\cite{McGregor}.
There, spatial spin splitting was described by using only a strong magnetic field gradient over the diffraction slits.
The approach given above delivers new insight into this problem, and couples the previous result to holographic vortex creation.
The exact details of such a magnetic aperture are left for future research, but the outline given here should provide a direction worthy \revision{of pursuit}.
}

\revision{
We will now investigate under what conditions the intrinsic relativistic spin coupling (detailed in the next section) might be exploited in the kinematical regime of a TEM.
}
\section{OAM Dirac spinors} \label{sec:Dirac_spinor}
\subsection{Mathematical solutions}

In order to further understand the role of spin and thus the \revision{total angular momentum} of electron vortex states, the solutions of the cylindrical Dirac equation are considered.
The underlying foundation of the question whether $L_z$ or $J_z$ is the most important variable, can be answered quantitatively by inspecting the fully relativistically correct cylindrical solutions to the free Dirac equation ($\gamma^\mu$ are the Dirac matrices, $\Psi$ is a 4-component Dirac spinor):
\begin{equation}
\left(\i\gamma^\mu\partial_\mu - m\right)\Psi = 0.
\end{equation}
This equation admits cylindrically symmetric solutions of the form:
\begin{align} \label{eq:dirac_cylindrical_spinor}
\Psi_{n,s}&(r,\varphi,z) = \e^{\i \left( k_z z - E t\right)} \times \notag \\
&
\begin{pmatrix}
\e^{\i n\varphi} J_n (k_\perp r) \\
s \e^{\i(n+1)\varphi} J_{n+1} (k_\perp r) \\
\frac{k_z - \i s k_\perp}{E+m} \e^{\i n\varphi} J_n (k_\perp r) \\
-s \frac{k_z - i s k_\perp}{E+m} \e^{\i(n+1)\varphi} J_{n+1} (k_\perp r)
\end{pmatrix}.
\end{align}
These are eigenstates of $E$, $k_z$, $J_z$, $p_\perp$, and the transverse helicity defined by $h_\perp=\gamma^5\gamma^3\frac{\bs \Sigma \cdot \bs p_\perp}{|\bs p_\perp|}$, which takes the values $\pm1$.
Taking a similar linear combination as the one that relates the Pauli spinors in equations \eqref{eq:pauli_vortex_spin} and \eqref{eq:pauli_vortex_j}, one can construct from \eqref{eq:dirac_cylindrical_spinor} only an approximate $L_z$ eigenstate:
\begin{subequations} \label{eq:semi_Lz}
  \begin{align} \label{eq:semi_Lz_plus}
  \Psi_n^{(+)} &= \frac{1}{2} \left( \Psi_{n,s=+1} + \Psi_{n,s=-1} \right) \notag \\
  &\propto 
  \begin{pmatrix}
  \e^{\i n\varphi} J_n(k_\perp r) \\
  0 \\
  \frac{k_z}{E+m} \e^{\i n\varphi} J_n(k_\perp r) \\
  \frac{\i k_\perp}{E+m} \e^{\i(n+1)\varphi} J_{n+1}(k_\perp r)
  \end{pmatrix}
  \end{align}
  \begin{align} \label{eq:semi_Lz_min}
  \Psi_{n-1}^{(-)} &= \frac{1}{2} \left( \Psi_{n-1,s=+1} - \Psi_{n-1,s=-1} \right) \notag \\
  &\propto
  \begin{pmatrix}
  0 \\
  \e^{\i n\varphi} J_n(k_\perp r) \\
  \frac{-\i k_\perp}{E+m} \e^{\i(n-1)\varphi} J_{n-1} (k_\perp r) \\
  \frac{-k_z}{E+m} \e^{\i n\varphi} J_n(k_\perp r)
  \end{pmatrix}.
  \end{align}
\end{subequations}
The solutions in equation \eqref{eq:semi_Lz} bear the closest resemblance to the Pauli spinors discussed above (see \eqref{eq:pauli_vortex_spin}).
These solutions are still $J_z$ eigenstates with eigenvalue $n\pm1/2$, and become $L_z=n$ and $\Sigma_z$ (spin) eigenstates only in the paraxial approximation: $k_\perp \ll E+m$.
Note that both of these solutions carry a different unapproximated eigenvalue of $J_z$, although their approximate $L_z$ eigenvalue is equal.
It is as though the paraxial limit removes the coupling of the electronic wave function to its spin in this field-free case.
Bliokh \textit{et al.}~\cite{Bliokh} have previously presented an alternative form of \eqref{eq:semi_Lz}, but the form here better captures the kinematical quantities that are relevant in a practical TEM experiment.
\revision{Note that these are basis functions; a real vortex beam's transverse wave function is generally a superposition of many $k_\perp$-states\footnote{\revision{Several differing expressions have been given in literature depending on the exact form of the hologram and input beam~\cite{Janicijevic, Heckenberg_fraunhofer, Vasara}.
Other ways of producing vortex beams like mode transformation\cite{Schattschneider_mode_conversion} will again give different expressions for the transverse wave function.}}.
}

The quantitative analysis of the relativistic spin coupling these solutions contain is presented below.

\subsection{Relativistic spin coupling in electron vortex beam states}

As previously illuminated by Bliokh \textit{et al.}~\cite{Bliokh}, there are two special states which, in principle, permit us to directly observe a relativistic effect.
There, it is shown that for the states $\Psi_{-1}^{(+)}$ and $\Psi_1^{(-)}$ (see \eqref{eq:semi_Lz}), there is a significant central contribution to the density $\rho$.
\revision{

We will first show that the states described in~\cite{Bliokh} and the ones in \eqref{eq:semi_Lz} are indeed exactly the same up to a trivial normalization factor.
The equivalence follows directly from Fig. 1a from~\cite{Bliokh}: the understanding that here $k_z$ and $k_\perp$ equal ref.~\cite{Bliokh}'s $k \cos{\theta_0}$ and $k \sin{\theta_0}$ respectively, with $k^2=E^2-m^2=k_z^2+k_\perp^2$.
Dividing ref.~\cite{Bliokh}'s expression by $\sqrt{1+\frac{m}{E}}$, it is immediate that the components' prefactors are identical to those in \eqref{eq:semi_Lz} here, with $\hbar=c=1$.
The value $\theta_0 = \pi/4$ means $k_\perp=k_z$, which is unrealistic in any current setup.
}

\begin{figure}
  \centering
  \includegraphics[width=\linewidth]{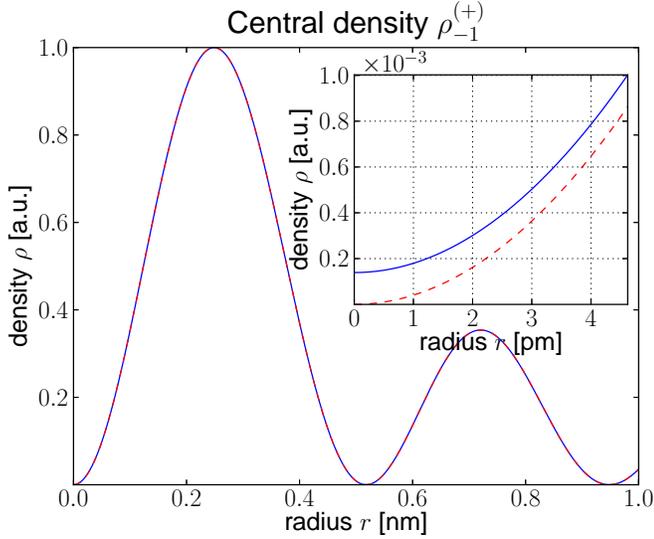}
  \caption{Central density distribution (\ref{eq:dirac_density}) using the parameters mentioned in the text. The blue (solid) line is given by \eqref{eq:dirac_density} and the red (dashed)  line is only the first term of that equation. \revision{The inset zooms in on the central region.} \label{fig:central_density}}
\end{figure}

Below is shown however that this effect is extremely small for typical \revision{TEM} parameters.
The zeroth order Bessel functions are the only ones that are nonzero at the origin (\textit{i.e.} the center of the beam).
This gives the states $\Psi_{-1}^{(+)}$ and $\Psi_1^{(-)}$ (see equation \eqref{eq:semi_Lz}) a nonzero central density:
\begin{align} \label{eq:dirac_density}
\rho_{-1}^{(+)}(r) = \rho_1^{(-)} (r) = &\left(1+\frac{k_z^2}{(E+m)^2}\right)J_{1}^2(k_\perp r) \notag \\
&+ \frac{k_\perp^2}{(E+m)^2} J_0(k_\perp r).
\end{align}
The third term results in a non-zero density at $r=0$, proportional to $\frac{k_\perp^2}{(E+m)^2}$.
Approximating the radius of the vortex \revision{$R$} by only looking at the maximum of $J_1^2$, one can estimate $k_\perp$ (up to $\mathcal{O}\left((k_\perp r)^4\right)$) for a single-$k_\perp$ state:
\begin{equation} \label{eq:size_kperp}
k_\perp \; (\text{in keV}) \approx \frac{0.37\hdots}{R \; (\text{in nm})}.
\end{equation}
In state of the art \revision{TEM} experiments, one can achieve a focused electron vortex~\cite{Verbeeck} of $R\approx 0.5$\;\AA, and one has $k_\perp\approx 7.4$\;keV.
For a $200$\;keV beam, this results in a central contribution to the density of about $3.7\times 10^{-5}$.
This figure needs to be compared with the unity sized contribution of the other terms in \eqref{eq:dirac_density}.
Fig. \ref{fig:central_density} shows a comparison of the first term of \eqref{eq:dirac_density} with the full expression for the aforementioned values.
It should be compared with fig. 2 in~\cite{Bliokh}.
The relativistic effect for these states in this parameter regime are much smaller than the differences shown in that article.
By increasing the focus and thus by further shrinking the vortex, one can in principle increase the magnitude of the effect, because $k_\perp$ increases with decreasing vortex size according to \eqref{eq:size_kperp}.
Though as one increases $k_\perp$, the relevant area of the central density will shrink as well.
One can easily estimate the radius inside which the zeroth order contribution is larger than the first order in \eqref{eq:dirac_density}:
\begin{equation}
r_C \approx \frac{0.24\hdots}{\left(k_z^2+m\left(m+\sqrt{k_z^2+m^2}\right)\right)^{1/4}} \sqrt{R}+\hdots,
\end{equation}
where $R$ is in nm and the kinematic quantities in keV (natural units).
For our values of the parameters $k_\perp$ and $k_z$, one has \revision{$r_C\approx 1.8$\;pm}, which is of the order of the electron wavelength.
The area inside the critical radius is thus of the order of 50\;pm$^2$, in which the intensity is equal to approximately $10^{-5}$ the intensity at the vortex maximum.
For a spin unpolarized beam the oppositely polarized vortex state (see \eqref{eq:semi_Lz}) is also present, which adds another unwanted unity sized background contribution which has to be discriminated upon measurement.
The extremely small size of these effects allows us to say with certainty that pure relativistic effect of free-space electron vortices is unobservable for the conditions prevalent in an electron microscope.
The smallness of these effects will translate into the fact that the Pauli description \revision{suffices}, and effectively $L_z$ can be maintained as a good quantum number even though this is not the case in the Dirac description.

\section{Conclusions}

\revision{In this paper the properties of electron vortex states have been investigated in the experimental setting of current electron microscopes.}
The question of the right quantum numbers to qualify electron vortex states in \revision{field-free conditions} is discussed.

\revision{
The usage of a holographic mask to poduce $J_z$ eigenstates was investigated.
A holographic fork aperture works spatially, and localized magnetic fields will need to be added for spin to come into the picture.
The introduction of an electromagnetic field will couple the two components, and we expect the $J_z$ eigenstates discussed here to become important.
One can thus not fabricate a $J_z$ holographic mask without magnetizing it.
This extra degree of manipulation could in itself lead to transverse spin-splitting as in~\cite{McGregor}, although it is not clear how the transverse Stern-Gerlach effect calculated there will couple to the vortex formation by a holographic mask simultaneously.
A combined magnetized and forked pattern seems like the most achievable solution, once the necessary magnetization conditions can be experimentally achieved.
}

In an attempt to quantify the relativistic spin coupling effects as previously illuminated by Bliokh \textit{et al.}~\cite{Bliokh}, an alternative form of the Dirac spinors representing approximate $L_z$ electron vortex eigenstates is presented and their connection to the transverse helicity eigenstates of the Dirac equation is shown.
These were used to calculate the size of the contribution of the relativistic description to the central density present in specific states.
It was found that for realistic conditions in modern electron microscopes, \revision{the relativistic spin coupling effects} are very difficult, if not impossible, to observe: the relative intensity of this central density compared to the maximum of the full wave lies around $10^{-5}$, and the size of the area in which the effect dominates is of the order of the electron wavelength.
The current experimental apparatus does not lead to sufficiently large relativistic \revision{spin coupling effects, and thus} these cannot be measured.

In principle, one can repeat the calculations in this paper for neutrons, where perhaps the different interactions and relevant parameters can cause the effects described above to become observable.
Neutron vortex states have not been presented as far as the authors are aware, but might bring the rich properties of phase singular beams to neutron experiments.




\acknowledgments
This research was supported by an FWO PhD fellowship grant (Aspirant Fonds Wetenschappelijk Onderzoek - Vlaanderen).
The authors acknowledge financial support from the European Union under the Seventh Framework Program under a contract for an Integrated Infrastructure Initiative. Reference No. 312483-ESTEEM2.
J. Verbeeck acknowledges funding from the European Research Council under the 7th Framework Program (FP7), ERC grant N°246791 - COUNTATOMS and ERC Starting Grant 278510 VORTEX.
\revision{The authors would like to thank the referees for their constructive comments.}

\bibliographystyle{eplbib}
\bibliography{spin_vortex}

%
%
%
%

\end{document}